\documentclass[preprint]{elsarticle}

\usepackage{lineno,hyperref}
\modulolinenumbers[5]

\usepackage{physics}
\usepackage{siunitx}
\usepackage{amsmath}
\usepackage{booktabs}

\journal{Journal of Luminescence}

\begin{document}

\begin{frontmatter}

\title{A Raman Heterodyne Study of the Hyperfine Interaction of the Optically-Excited State $^5$D$_0$ of Eu$^{3+}$:Y$_2$SiO$_5$}


\author{Yu Ma\corref{}}
\author{Zong-Quan Zhou\corref{mycorrespondingauthor1}}
\cortext[mycorrespondingauthor1]{Corresponding author}
\ead{zq\_zhou@ustc.edu.cn}
\author{Chao Liu\corref{}}
\author{Yong-Jian Han\corref{}}
\author{Tian-Shu Yang\corref{}}
\author{Tao Tu\corref{}}
\author{Yi-Xin Xiao\corref{}}
\author{Peng-Jun Liang\corref{}}
\author{Pei-Yun Li\corref{}}
\author{Yi-Lin Hua\corref{}}
\author{Xiao Liu\corref{}}
\author{Zong-Feng Li\corref{}}
\author{Jun Hu\corref{}}
\author{Xue Li\corref{}}
\author{Chuan-Feng Li\corref{mycorrespondingauthor2}}
\cortext[mycorrespondingauthor2]{Corresponding author}
\ead{cfli@ustc.edu.cn}
\author{Guang-Can Guo\corref{}}

\address{CAS Key Laboratory of Quantum Information, University of Science and Technology of China, Hefei, 230026, P.~R.~China}
\address{Synergetic Innovation Center of Quantum Information and Quantum Physics, University of Science and Technology of China, Hefei, 230026, P.~R.~China}

%


\begin{abstract}
The ground-state spin coherence time of $^{151}$Eu$^{3+}$ in Y$_2$SiO$_5$ crystal in a critical magnetic field was extended to six hours in a recent work [\textit{Nature} \textbf{517}, 177 (2015)], which paved the way for constructing quantum memory with long storage time. In order to select a three-level system for quantum memory applications, information about the excited-state energy level structures is required for optical pumping. In this work, we experimentally characterize the hyperfine interaction of the optically-excited state $^5$D$_0$ using Raman heterodyne detection of nuclear magnetic resonance (NMR). The NMR spectra collected in 201 magnetic fields are well fitted. The results can be used to predict the energy level structures in any given magnetic field, thus enabling the design of optical pumping and three-level quantum memory in that field.
\end{abstract}

\begin{keyword}
Europium-doped yttrium orthosilicate \sep Rare-earth-ion-doped crystal \sep Excited-state hyperfine structure \sep Raman heterodyne spectroscopy
\end{keyword}

\end{frontmatter}


\section{Introduction}

As an essential component in quantum repeater scheme \cite{repeater}, quantum memory is the core element for the construction of large-scale quantum network. Experiments have been performed in many systems such as atomic gases \cite{atomensembleqm,atomensembleqm1,atomensembleqm2}, single atom \cite{singleatomqm}, molecular gases \cite{moleensembleqm} and rare-earth-ion-doped crystals (REICs).

It is noteworthy that REICs have drawn great interest over the last decade, and proven to be a promising candidate for quantum memory with high efficiency \cite{efficientqm}, high fidelity \cite{hifiqm,oamentangle}, broadband operation \cite{broadbandqm} and long storage time \cite{longlifeqm}. Among all kinds of REICs, europium-doped yttrium orthosilicate (Eu$^{3+}$:Y$_2$SiO$_5$) is considered suitable for the realization of long-term memory. Extremely long spin relaxation time (22 days) in the ground state was observed \cite{EuYSOspectra} and the spin coherence time was extended to 6 hours in Eu$^{3+}$:Y$_2$SiO$_5$ using a carefully aligned magnetic field and dynamical decoupling \cite{6hour}. The unprecedented long coherence time could enable the construction of portable quantum hard drives for long-distance entanglement distribution.

The knowledge of energy level structures is the prerequisite for quantum memory protocols \cite{afcexp,cribexp,eitexp}. Appropriate optical pumping based on known structures can well isolate one class of ions within the inhomogeneous broadening \cite{SHBNdYVO}. In order to obtain the information of the hyperfine structures of REICs, two alternative approaches, spectral hole burning \cite{REbookMacSHB} and Raman heterodyne detection of NMR \cite{ramanprb,ramanprl,groundeu,groundpr}, are employed in previous works. Spectral hole burning is a high-resolution technique that has been used to study electronic transitions for a wide range of materials including REICs. Raman heterodyne method enables the characterization of hyperfine interactions with high signal-to-noise ratio (SNR) in such systems \cite{ramanprb,ramanprl}. Conventional NMR and nuclear quadrupole resonance (NQR) of Eu$^{3+}$ in solids have not been successful due to its small nuclear quadrupole moment \cite{REbookLiu}.

In this work, we present the measurements and fitting results of the excited-state hyperfine spectra of $^{151}$Eu$^{3+}$ in Y$_2$SiO$_5$ at crystallographic site 1, using Raman heterodyne detection of NMR. The spectra of the excited state $^5$D$_0$ are collected in a vector magnetic field of 800 Gauss. The orientation of the field is rotated in a spherical pattern and 201 NMR spectra are obtained in corresponding fields, based on which the effective spin Hamiltonians are determined with simulated annealing algorithm for two magnetically inequivalent subsites.

A shortcoming of Raman heterodyne method is that it only concerns the hyperfine transitions within the ground- or excited-state manifold and transitions of ions at two subsites are mixed up in the NMR spectra, such that it cannot tell to which subsite the ground- and excited-state hyperfine transitions belong. To eliminate the ambiguity, we perform a spectral hole burning experiment in a carefully chosen magnetic field by observing the anti-holes, from which the level structures for each subsite can be deduced.


\section{\label{theory}Theory}
The theory for hyperfine interactions of REICs has been investigated and elaborated in previous publications \cite{groundeu,groundpr,REbookMacHam,SHBEuCl3}. Here we give a brief review.

The Hamiltonian for 4f electron and nucleus of rare-earth ion has the form
\begin{equation}
H = H_{FI} + H_{CF} + H_{HF} + H_{Q} + H_{z} + H_{Z}.
\end{equation}
These six terms represent the free ion, crystal field, hyperfine, nuclear quadrupole, electronic Zeeman and nuclear Zeeman Hamiltonians, respectively. The free ion and crystal field terms determine the electronic energy levels and optical transitions. Due to the fact that there is no net electron spin or orbital angular momentum in Eu$^{3+}$:Y$_2$SiO$_5$, the other four terms are small enough to be treated as a perturbation for the electronic state. Using second order perturbation theory and discarding the perturbation term that does not influence the hyperfine energy level, we can obtain the effective spin Hamiltonian \cite{groundeu,groundpr}
\begin{equation}
H = B\cdot M\cdot I + I\cdot Q\cdot I,
\end{equation}
where $B$ is the external magnetic field. $M$ and $Q$ are the effective Zeeman and quadrupole tensors for rare-earth ion. $I$ is the nuclear spin operator, which for spin-5/2 nuclei like Eu$^{3+}$ can be parameterized as three $6\times6$ matrices. The parameterization is presented in \ref{fitting}.

Y$_2$SiO$_5$ has a symmetry which can be described by $C_{2h}^{6}$ space group. Yttriums occupy two sites in the crystal. Eu$^{3+}$ as a dopant substituting yttrium also occupies two sites, 1 and 2, exhibiting two inhomogeneously broadened absorption features at 580.039 nm and 580.211 nm respectively \cite{EuYSOwavelength2013}. Each site can be divided into two magnetically inequivalent subsites, related to each other by the crystal $C_2$ axis. Here we measure the excited-state hyperfine structure of $^{151}$Eu$^{3+}$ at site 1. This particular choice of isotope and crystallographic site is made because it has demonstrated the longest spin coherence time so far \cite{6hour}.

Due to the $C_2$ symmetry, two subsites of Eu$^{3+}$ should be described by different $M$ and $Q$ tensors. Each can be transformed into the other by a $\pi$-rotation operation about the $C_2$ axis.

\section{\label{exp}Experiment}

\subsection{\label{Raman}Raman heterodyne detection}
Raman heterodyne detection of NMR, as an RF-optical double-resonance technique based on the coherent Raman effect, is a useful tool for observing NMR \cite{ramanprb,ramanprl}. Measurements of the ground-state hyperfine structure $^7$F$_0$ of $^{151}$Eu$^{3+}$ at site 1 in Y$_2$SiO$_5$ were covered elsewhere \cite{groundeu}, which provide a stepping stone towards hour-long spin coherence time. Here we present the characterization of the hyperfine interaction of the excited state $^5$D$_0$ of $^{151}$Eu$^{3+}$ at the same site using a similar approach.

\begin{figure}[htbp]
\centering
\includegraphics[width=\textwidth,trim=0 0 0 0]{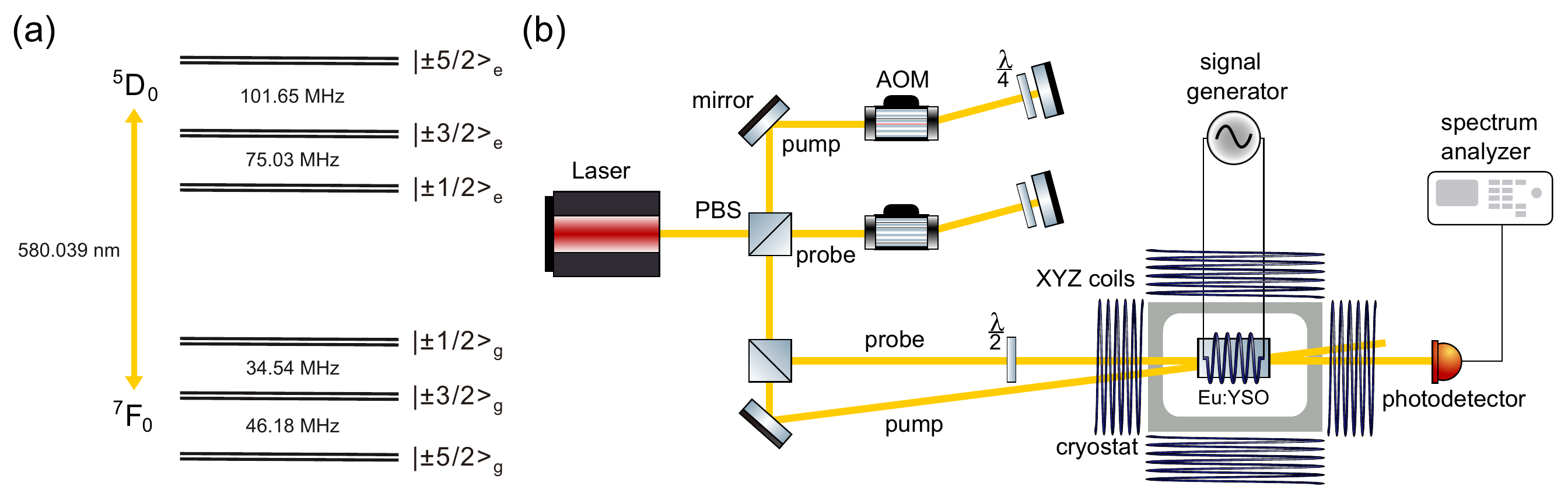}
\caption{\label{setup} (Color online) (a) The hyperfine structure of the ground state $^7$F$_0$ and excited state $^5$D$_0$ of $^{151}$Eu$^{3+}$ at site 1 of Y$_2$SiO$_5$ in zero magnetic field. (b) Experimental setup for Raman-heterodyne-detected NMR of Eu$^{3+}$:Y$_2$SiO$_5$. The laser is split into two paths by a polarization beam splitter (PBS). The frequency of the light is modulated with an acousto-optic modulator (AOM) in a double-pass configuration. The polarization of the light is controlled by the half wave plate ($\lambda/2$) and quarter wave plate ($\lambda/4$). The probe laser is incident on the crystal placed inside a cryostat. The half wave plate in front of the cryostat rotates the polarization of the beam to the direction of strongest absorption. An RF field generated by a 7-turn coil wrapped around the sample drives the hyperfine transitions. The pump laser is employed to create a population difference between two hyperfine states. The photodetector captures the beat signal produced by the transmitted light and the scattered light. The spectra are recorded with an analyzer.}
\end{figure}

The hyperfine energy level diagram in zero magnetic field is shown in Fig. \ref{setup}(a), which was determined by spectral hole burning experiment in previous works \cite{EuYSOwavelength1991,EuYSOnonlinear}. The experimental setup is depicted in Fig. \ref{setup}(b). A sample of 0.1 atomic (at.)\% Eu$^{3+}$:Y$_2$SiO$_5$ is used for the experiment. The crystal is cut along the three crystal axes $D_1$, $D_2$ and $b$ with a size of $4\times3\times10$ mm. The sample is placed inside a cryostat and cooled to approximately 3.5 K for spectrum measurements. The XYZ coils outside the cryostat are used to provide a magnetic field. An RF field is produced by a 7-turn coil wrapped around the sample along the $C_2$ axis. Two laser beams from a frequency-doubled semiconductor laser tuned to be resonant with the $^7$F$_0$ $\rightarrow$ $^5$D$_0$ transition with a linewidth well below 10 kHz are incident on the sample. The probe beam propagates along the crystal $C_2$ axis with a power of 2.3 milliwatts and a diameter of 50 $\mu$m inside the crystal. The transmitted light which carries the Raman heterodyne signal is detected with a photodetector. The pump beam has a diameter of 120 $\mu$m, with a power of 7 milliwatts. The two beams overlap inside the crystal, with a small angle between their propagating directions.

The Raman heterodyne signal is generated when the RF field is resonant with the hyperfine transition either in the excited- or ground-state manifold. The signal stems from a beat between the transmitted probe laser and the Raman scattered light.

In our experiment, we find that an appropriate pump laser can significantly increase the Raman heterodyne signal. For example, the frequency of the pump laser is swept around the $\ket{\pm1/2}_g\leftrightarrow\ket{\pm3/2}_e$ for the measurements of $\ket{\pm1/2}_e\leftrightarrow\ket{\pm3/2}_e$. The SNR with the pump laser is significantly higher than that without it. There are two reasons that may account for this phenomenon. One is the linear relationship between Raman heterodyne signal and the population difference between the resonant levels \cite{ramanprb}. A pump laser can pump away the ions on the neighbor levels thus increasing the population difference. The other reason is similar to what is mentioned in Ref. \cite{groundeu}, where J. J. Longdell \textit{et al.} did not find the signal without a pump light. Here, if the signal laser is incident without pump, all of the ground states contribute to the signal, while a pump laser can break the symmetry, thus increasing the signal.

\subsection{\label{spectra}Spectrum measurements}

In order to obtain the information about hyperfine structures of Eu$^{3+}$ in magnetic fields, a DC magnetic field of 800 Gauss is applied to the crystal to split the degenerate hyperfine energy levels. We follow the similar approach carried out for the ground state $^7$F$_0$ \cite{groundeu,groundpr}, rotating the direction of the magnetic field in a spherical pattern
\begin{equation}
\left\{
\begin{aligned}
B_x &= B_0\sqrt{1-t^2}cos(6\pi t)\\
B_y &= B_0\sqrt{1-t^2}sin(6\pi t)\\
B_z &= B_0t
\end{aligned}
\right.
\end{equation}
in which $t\in[-1,1]$.

The formula is given in laboratory coordinate system, with three axes corresponding to the XYZ coils. The crystal is placed in the cryostat by hand, such that the crystal coordinate system $[D_1,D_2,b]$ has a small misalignment with the laboratory system. This misalignment does not bring extra error to the measurements because it can be fitted in the fitting procedure.

\begin{figure}[htbp]
\centering
\includegraphics[width=\textwidth,trim=0 0 0 0]{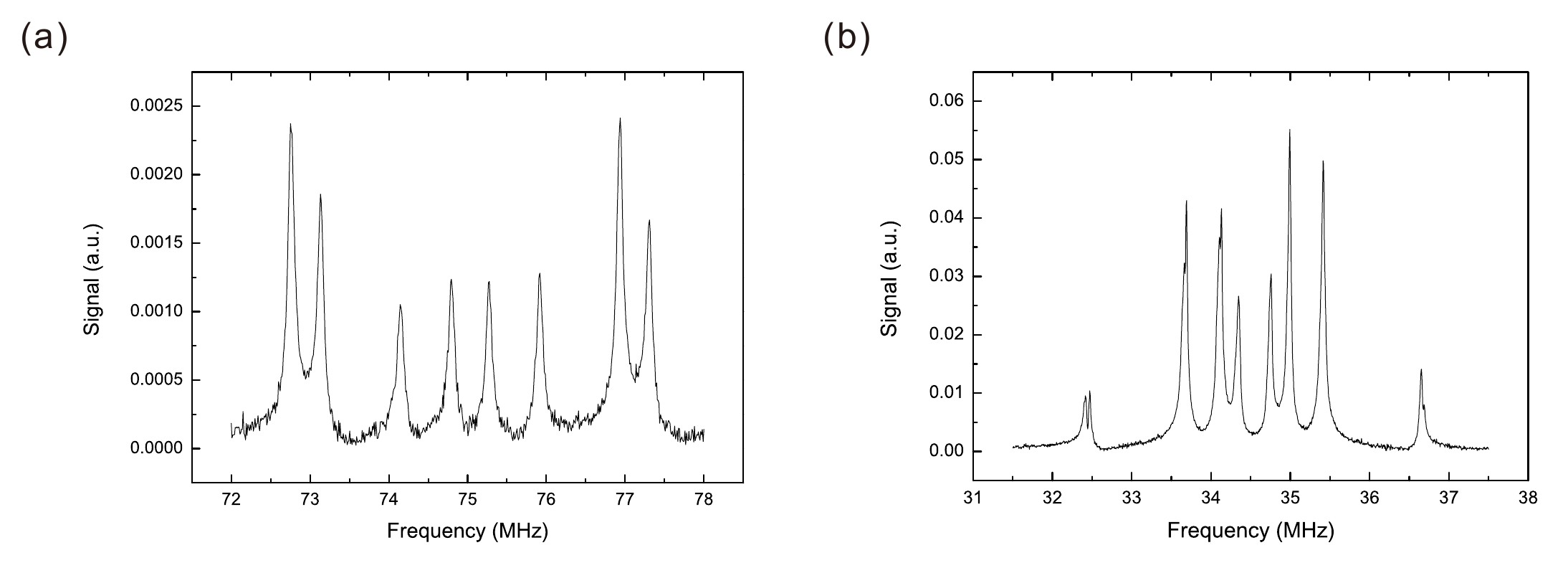}
\caption{\label{RHsignal} Two typical Raman-heterodyne spectra for the excited (a) and ground (b) state collected in magnetic field corresponding to $t=-0.32$ and $t=-0.45$, respectively. Eight peaks can be clearly located with high SNR for both the excited and ground state. Unexpected tiny splittings are observed for $\ket{\pm1/2}_g\leftrightarrow\ket{\pm3/2}_g$ transition around 34.5 MHz. The mechanism of these splittings requires more investigations.}
\end{figure}

The spectrum measurements are carried out around 75 MHz and 101.7 MHz for two excited-state hyperfine transitions $\ket{\pm1/2}_e\leftrightarrow\ket{\pm3/2}_e$ and $\ket{\pm3/2}_e\leftrightarrow\ket{\pm5/2}_e$ respectively in 201 magnetic fields, corresponding to $t\in[-1,1]$ uniformly divided into 201 points. Over 1000 averages are taken to reach a high SNR for each spectrum. The SNR of the observed spectra is sufficient for determining the parameters of the Hamiltonian.

In order to give a energy level diagram for both the ground and excited state in any given magnetic field, measurements are also performed around 34.5 MHz and 46.2 MHz for two ground-state hyperfine transitions $\ket{\pm1/2}_g\leftrightarrow\ket{\pm3/2}_g$ and $\ket{\pm3/2}_g\leftrightarrow\ket{\pm5/2}_g$.


Two typical spectra for the excited-state transition $\ket{\pm1/2}_e\leftrightarrow\ket{\pm3/2}_e$ and ground-state transition $\ket{\pm1/2}_g\leftrightarrow\ket{\pm3/2}_g$ in magnetic fields are shown in Fig. \ref{RHsignal}. The degeneracy of $\ket{\pm1/2}$ and $\ket{\pm3/2}$ state is removed in magnetic fields thus one NMR peak splits into four. Eight NMR peaks are detected in the experiment, four of which represent the hyperfine transitions of ions at one subsite of the two.

Unexpected splittings are observed for the $\ket{\pm1/2}_g\leftrightarrow\ket{\pm3/2}_g$ transition. Typical examples are the first three peaks and the last one shown in Fig. \ref{RHsignal} (b). The largest splitting we observed is approximately 100 kHz. We also perform the measurements for a 0.01 at.\% sample. These tiny splittings still appear although the linewidth of each peak is narrower for a lower doping concentration. Further studies are required to identify the mechanisms for these splittings. As a reasonable approximation, we take the average value of the positions of the split peaks as the hyperfine transition frequency for fitting procedure.

\begin{figure*}[htbp]

\includegraphics[width=\textwidth,trim=0 0 0 0]{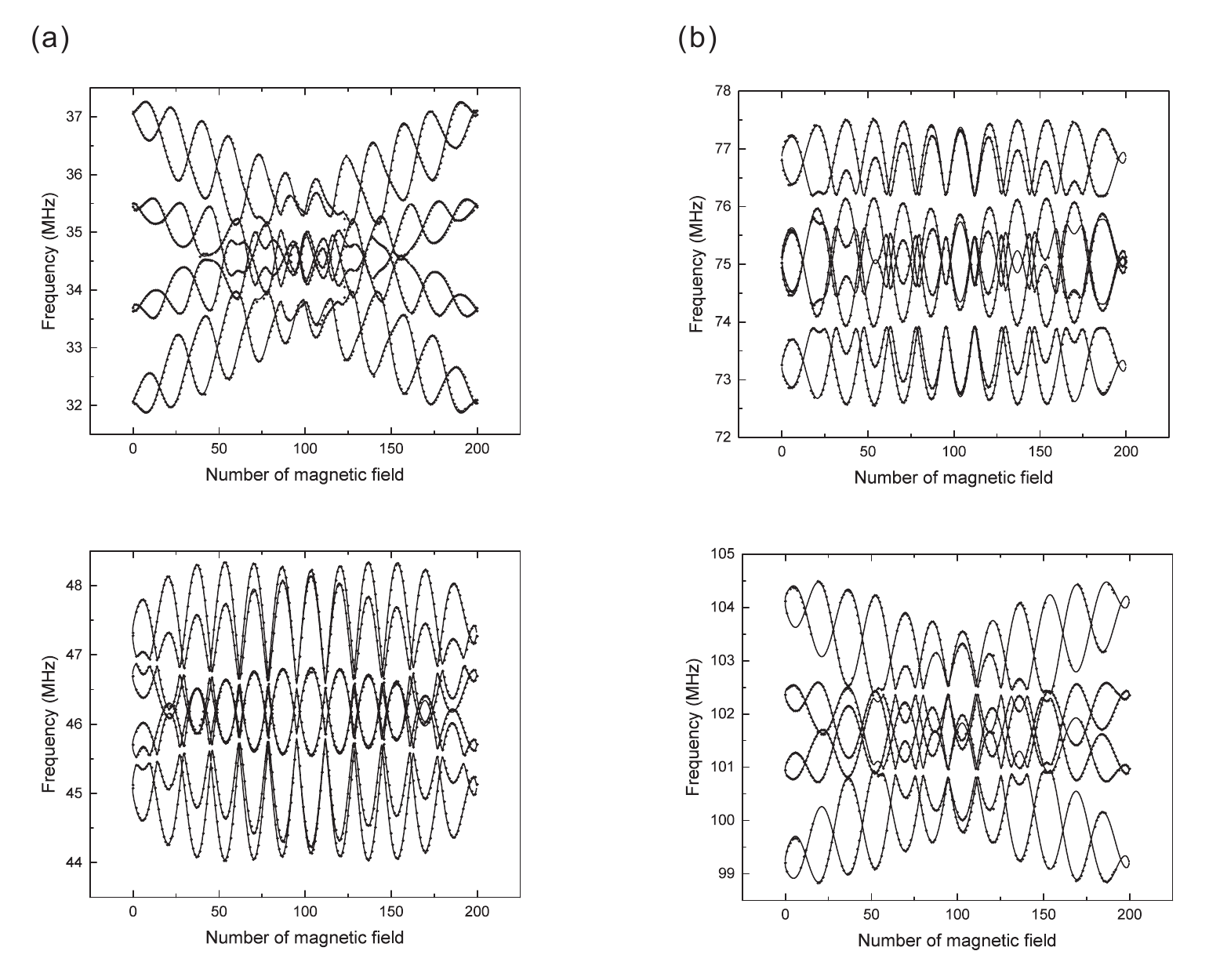}
\caption{\label{fitfig} The experimental data and fit curves for the ground-state (a) and excited-state (b)  hyperfine transitions in 201 magnetic fields. The dots represent the NMR peak positions detected in the experiment, and the solid lines are the fit curves of the experimental data.}
\end{figure*}

\section{\label{fitting}Results and discussion}

\subsection{\label{fitting}Fitting procedure}
The parameterization of the effective spin Hamiltonian of the $^5$D$_0$ state is the same as that used for the ground state \cite{groundeu}. Two parameters are required to determine the Q tensor in its principal axis system:
\begin{equation}
\left( {\begin{array}{*{20}{c}}
{ - E}&0&0\\
0&E&0\\
0&0&D
\end{array}} \right).
\end{equation}
Then the Q tensor in the lab coordinate system can be obtained by a rotation characterized by Euler angles, \textit{i. e.}:
\begin{equation}
{Q_1} = R({\alpha _Q},{\beta _Q},{\gamma _Q})\left( {\begin{array}{*{20}{c}}
{ - E}&0&0\\
0&E&0\\
0&0&D
\end{array}} \right){R^T}({\alpha _Q},{\beta _Q},{\gamma _Q}).
\end{equation}
The subscript 1 denotes one of the two magnetically inequivalent subsites. The rotation matix $R(\alpha,\beta,\gamma)$ can be written in the form:
\begin{equation}
\begin{array}{ll}
R(\alpha ,\beta ,\gamma ) =& \left( {\begin{array}{*{20}{c}}
{\cos \alpha }&{ - \sin \alpha }&0\\
{\sin \alpha }&{\cos \alpha }&0\\
0&0&1
\end{array}} \right) \cdot \left( {\begin{array}{*{20}{c}}
{\cos \beta }&0&{\sin \beta }\\
0&1&0\\
{ - \sin \beta }&0&{\cos \beta }
\end{array}} \right)\\
 &\cdot \left( {\begin{array}{*{20}{c}}
{\cos \gamma }&{ - \sin \gamma }&0\\
{\sin \gamma }&{\cos \gamma }&0\\
0&0&1
\end{array}} \right),
\end{array}
\end{equation}
where the subscript Q has been taken away since the rotation matrix has the same form for the M tensor:
\begin{equation}
{M_1} = R({\alpha _M},{\beta _M},{\gamma _M})\left( {\begin{array}{*{20}{c}}
{{g_1}}&0&0\\
0&{{g_2}}&0\\
0&0&{{g_3}}
\end{array}} \right){R^T}({\alpha _M},{\beta _M},{\gamma _M}).
\end{equation}

$M_2$ and $Q_2$ tensors for ions at another subsite can be easily obtained by a $\pi$-rotation operation on $M_1$ and $Q_1$ about the crystal $C_2$ axis. We assume an azimuthal angle $\theta$ and a polar angle $\phi$ for the direction of the $C_2$ axis because of the misalignment between the laboratory and crystal system mentioned in the last section. The $\pi$-rotation operation can be written in a matrix form \cite{c2angle}:
\begin{equation}
{R_\pi }(\theta_{C_2},\phi_{C_2} ) = I + 2J(\theta_{C_2} ,\phi_{C_2} ){J^T}(\theta_{C_2} ,\phi_{C_2} ),
\end{equation}
where $J(\theta ,\phi )$ is the skew symmetric operator:
\begin{equation}
J(\theta ,\phi ) = \left( {\begin{array}{*{20}{c}}
0&{ - \cos \theta }&{\sin \theta \sin \phi }\\
{\cos \theta }&0&{ - \sin \theta \cos \phi }\\
{ - \sin \theta \sin \phi }&{\sin \theta \cos \phi }&0
\end{array}} \right).
\end{equation}

The spin-5/2 operator can be written as three $6\times6$ matrices:
\begin{equation}
{I_x} = \dfrac{1}{2}\left( {\begin{array}{*{20}{c}}
0&\sqrt{5}&0&0&0&0\\
\sqrt{5}&0&2\sqrt{2}&0&0&0\\
0&2\sqrt{2}&0&3&0&0\\
0&0&3&0&2\sqrt{2}&0\\
0&0&0&2\sqrt{2}&0&\sqrt{5}\\
0&0&0&0&\sqrt{5}&0\\
\end{array}} \right),
\end{equation}

\begin{equation}
{I_y} = \dfrac{i}{2}\left( {\begin{array}{*{20}{c}}
0&-\sqrt{5}&0&0&0&0\\
\sqrt{5}&0&-2\sqrt{2}&0&0&0\\
0&2\sqrt{2}&0&-3&0&0\\
0&0&3&0&-2\sqrt{2}&0\\
0&0&0&2\sqrt{2}&0&-\sqrt{5}\\
0&0&0&0&\sqrt{5}&0\\
\end{array}} \right),
\end{equation}

\begin{equation}
{I_z} = \dfrac{1}{2}\left( {\begin{array}{*{20}{c}}
5&0&0&0&0&0\\
0&3&0&0&0&0\\
0&0&1&0&0&0\\
0&0&0&-1&0&0\\
0&0&0&0&-3&0\\
0&0&0&0&0&-5\\
\end{array}} \right).
\end{equation}

After the parameterization of the spin Hamiltonian, one can obtain the eigen-energy spectrum by solving the stationary Schr\"{o}dinger equation, as long as the parameters are assigned with certain values. For the fitting procedure, we first use a random set of values to parameterize the Hamiltonians. Then we minimize the error between the calculated spectra and the experimental ones by continuously optimizing the parameters.

The error minimization is carried out by computer using simulated annealing algorithm \cite{simuanneal}. The objective function is set to be the total absolute error of all the spectra of both the excited and ground state. In the fitting procedure, we find that the order of the energy levels is related to the signs of the diagonal elements of the $M$ and $Q$ tensor, \textit{i. e.} $g_1$, $g_2$, $g_3$, $E$ and $D$. In order to obtain the correct order which has been identified in zero field \cite{151EuYSOmemory}, we do not restrict those parameters to be positive as that in Ref. \cite{groundeu}.

The fitting results are shown in Fig. \ref{fitfig} and Tab. \ref{fitable}. Note that the excited and ground state share the same $\theta_{C_2}$ and $\phi_{C_2}$ since the orientation of the $C_2$ axis is fixed in our laboratory system, which is independent of which state we investigate. Also, the azimuth angle is small, which confirms that the crystal coordinate system is approximately aligned with the laboratory system. The final fitting error is reduced to roughly 14 kHz per peak on average, which is comparable with the inhomogeneous broadening. The absolute values of the diagonal elements of the $M$ and $Q$ tensors, \textit{i. e.} the $g$-factors, $E$ and $D$ of the ground state $^7$F$_0$, are close to those in Ref. \cite{groundeu}.

Uncertainties of the fit parameters are also presented in Fig. \ref{fitfig}. Since the results are given in the laboratory frame defined by the XYZ coils, the uncertainties do not include the calibration of the coils. Contribution of the calibration error to the uncertainties is expected to be less than 5\%.

We also note that it is necessary to give the direction of the $D_1$ and $D_2$ axis in our laboratory, so that one can transform the Hamiltonians into $[D_1,D_2,b]$ coordinate system, thus making the results more general. Since the direction of $b$ ($C_2$) axis can be fitted to the experiments above, the $D_1$-$D_2$ plane, perpendicular to the axis, is spontaneously determined. Eu$^{3+}$:Y$_2$SiO$_5$ presents the strongest absorption when the incident light is polarized along the $D_1$ axis \cite{EuYSOspectra,EuYSOabsorption}. As a reasonable approximation, the projection of the maximum absorption direction on the $D_1$-$D_2$ plane is considered to be parallel to the $D_1$ axis. The directions (in azimuth and elevation) of $D_1$ and $D_2$ are also listed in Tab. \ref{fitable}.

\begin{table}[tb]
\renewcommand{\arraystretch}{1.2}
\caption{Fitting results of the effective spin Hamiltonian parameters in the laboratory frame. The $^5$D$_0$ and $^7$F$_0$ state share the same $D_1$, $D_2$ and $b$ ($C_2$) axes, which are fixed in our laboratory system.}\label{fitable}
\centering
\begin{tabular}{SSSSS} \toprule
     & \multicolumn{2}{c}{Excited state $^5$D$_0$} & \multicolumn{2}{c}{Ground state $^7$F$_0$} \\
    {Parameter}   &  {Value}    & {Uncertainty}      &  {Value}    & {Uncertainty}\\ \midrule
    {$\alpha_M$(degrees)}       & -79.51    & 0.29      & -153.2297 & 0.0046\\
    {$\beta_M$(degrees)}        & 89.65     & 0.29      & -68.3     & 1.6\\
    {$\gamma_M$(degrees)}       & -68.670   & 0.072     & 10.113    & 0.098\\
    {$g_1$ (MHz/T)}             & 9.59      & 0.17      & 11.802    & 0.067\\
    {$g_2$ (MHz/T)}             & 9.444     & 0.047     & 4.7502    & 0.0035\\
    {$g_3$ (MHz/T)}             & 10.189    & 0.062     & -5.827    & 0.015\\
    {$\alpha_Q$ (degrees)}      & 35.4201   & 0.056     & -171.681  & 0.016\\
    {$\beta_Q$ (degrees)}       & -30.44    & 0.16      & -51.16    & 0.82\\
    {$\gamma_Q$ (degrees)}      & 1.3725    & 0.0003    & 49.4      & 2.9\\
    {$E$ (MHz)}                 & 5.8713    & 0.0002    & -2.7359   & 0.0001\\
    {$D$ (MHz)}                 & 27.18611  & 0.00001   & -12.3819  & 0.0001\\
    {$\theta_{C_2}$ (degrees)}  & 3.20      & 0.44      &           &   \\
    {$\phi_{C_2}$ (degrees)}    & 331.76    & 0.59      &           &   \\
    {$\theta_{D_1}$ (degrees)}  & 92.8220   &           &           &   \\
    {$\phi_{D_1}$ (degrees)}    & 0.0218    &           &           &   \\
    {$\theta_{D_2}$ (degrees)}  & 88.4860   &           &           &   \\
    {$\phi_{D_2}$ (degrees)}    & 89.9471   &           &           &   \\ \bottomrule
\end{tabular}
\end{table}

\subsection{\label{fitting}Subsite correspondence between the ground and excited state}
As described in Sec. \ref{theory}, Eu$^{3+}$ ions occupy two magnetically inequivalent subsites. Unfortunately, the fitted excited- and ground-state parameters given in Tab. \ref{fitable} may not refer to the same subsite only based on the Raman-heterodyne experiment above. In other words, since there are two $C_2$-symmetric subsites thus two $C_2$-related Hamiltonians for each state, one of the two ground-state Hamiltonians can be either related to one of the two excited-state Hamiltonians or the other. Raman-heterodyne-detected NMR method only concerns hyperfine transitions within an electronic state and does not provide any information about how the excited state is related to the ground state.

To eliminate the ambiguity, spectral hole burning technique is a useful tool to simultaneously investigate the ground- and excited-state hyperfine structures. Unlike Raman heterodyne method, hole-burning spectra contain information about hyperfine structures of both the ground and excited state. Distances between the holes (anti-holes) and the central hole reflect the hyperfine splittings in the ground and excited state \cite{SHBNdYVO,SHB,SHBEuYSO}, such that we can deduce the hyperfine structures from the hole-burning spectra.

However, spectral hole burning in a magnetic field is more complicated than that in zero field because the degeneracy of the Zeeman levels is removed. Three levels in zero field now split into six. Six ground levels combined with six excited levels lead to $6\times6=36$ classes of ions despite of the other $C_2$-symmetric subsite, while there only exist $3\times3=9$ classes in zero field. More energy levels create even more holes and anti-holes, which overlap with each other thus making the hole-burning spectra unrecognizable. To simplify the problem, it is necessary to use optical pumping to prepare a proper system with less upper and lower levels for the hole burning process \cite{SHB,SHBEuYSO}.

As a matter of fact, in our experiment there is no need to resolve all the holes and anti-holes. Therefore we designed a spectral hole burning experiment in a carefully chosen magnetic field which can isolate an anti-hole far from the central hole. This is done by calculating the possible positions of the holes and anti-holes using the parameters listed in Tab. \ref{fitable}. The calculations predict a characteristic anti-hole 3.45 MHz away from the central hole in a magnetic field of 600 Gauss in the direction of [0.6551,0.6976,-0.2900] in the laboratory system (and [0.6407,0.7053,-0.3034] in the $[D_1,D_2,b]$), while no holes or anti-holes are predicted beyond 3 MHz if conducting a $C_2$ rotation about the $b$ axis on the excited-state Hamiltonian.

The optical pumping sequences can be described in three steps as follows. We first employ three pump lasers, the frequency of which is swept over 8 MHz to address ions that is resonant with these three lasers, while other ions are pumped away. This part of the sequence, called class cleaning, is necessary to select only a subset of ions within the inhomogeneous broadening to prevent too many holes and anti-holes from appearing in the hole-burning spectrum. After class cleaning, the system is initialized into the $\ket{\pm1/2}_g$ state for hole burning by emptying another two ground levels. At last, a weak but sufficiently long pulse resonant with $\ket{\pm1/2}_g\leftrightarrow\ket{\pm1/2}_e$ is sent into the sample to burn a hole of considerable depth. Then we attenuate the signal laser pulse to single-photon level and sweep the frequency around the central hole to read out the spectral structures.

\begin{figure}[tb]
\centering
\includegraphics[width=\textwidth]{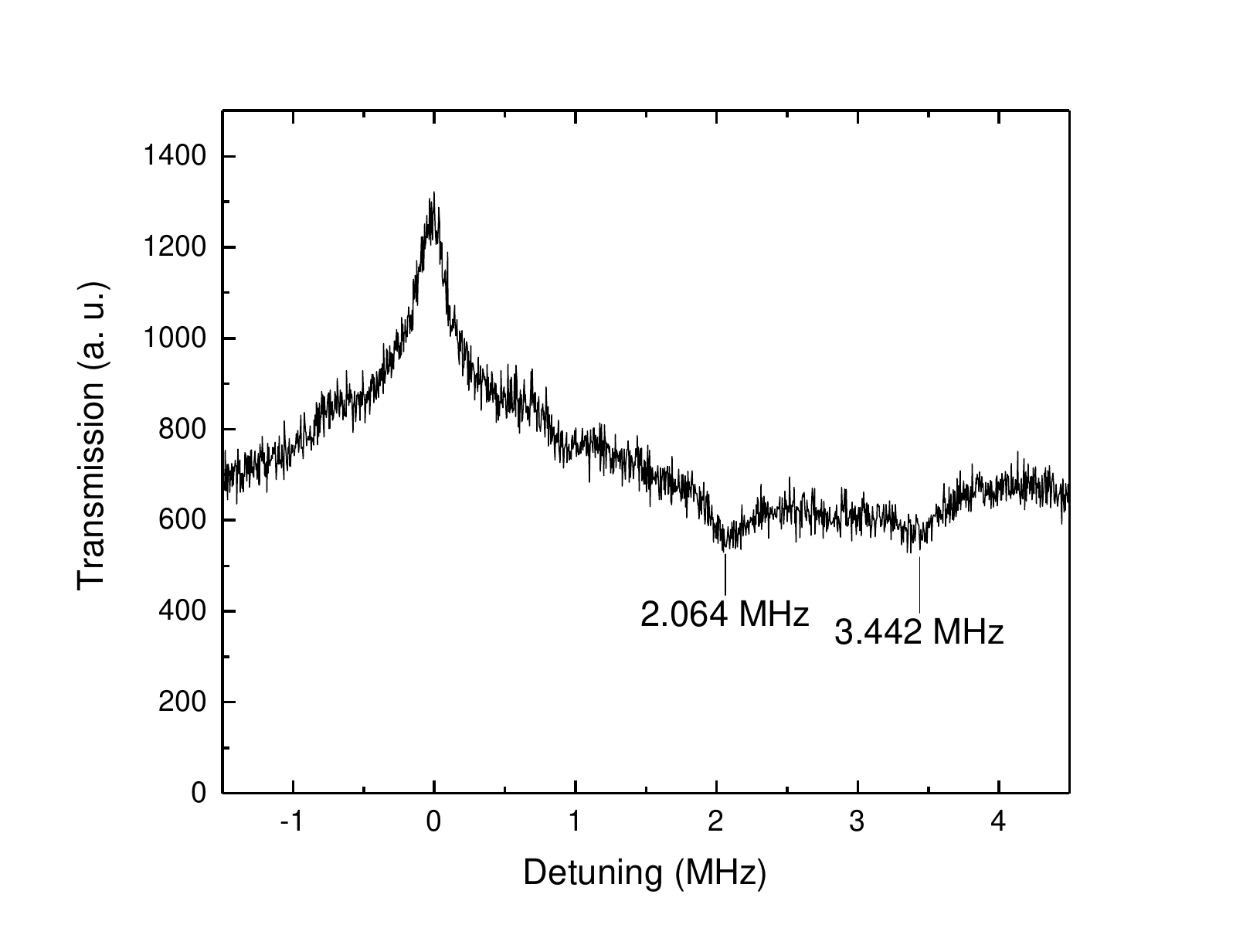}
\caption{\label{SHB}Hole-burning spectrum in a magnetic field of 600 Gauss in the direction of [0.6551,0.6976,-0.2900]. Two anti-holes can be resolved, which are consistent with the energy level structure predicted using the Hamiltonian parameters in Tab. \ref{fitable}. Other holes and anti-holes cannot be well located due to the limited SNR. The SNR is not as large as expected because the holes (and anti-holes) are broadened by the severe vibration caused by the cold head of the cryocooler.}
\end{figure}

The hole-burning spectrum is shown in Fig. \ref{SHB}. The existence of the anti-hole at 3.442 MHz verifies that the excited- and ground-state parameters listed in Tab. \ref{fitable} belong to the same subsite.

\subsection{\label{app}Applications}
\begin{figure}[tb]
\centering
\includegraphics[width=\textwidth]{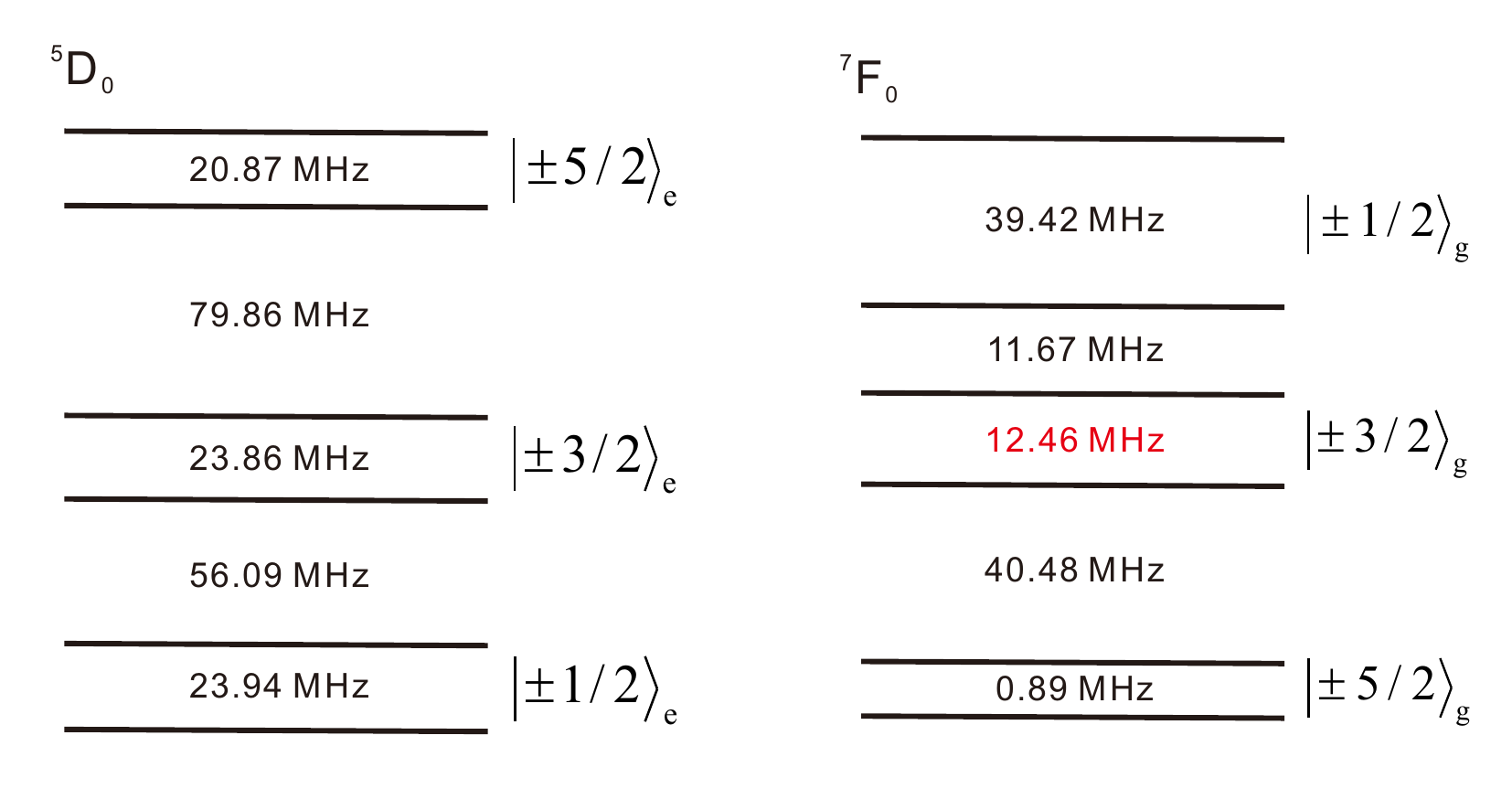}
\caption{\label{structure} (Color online) Hyperfine structure of the excited state $^5$D$_0$ and ground state $^7$F$_0$ of $^{151}$Eu$^{3+}$ at site 1 in Y$_2$SiO$_5$ in a ZEFOZ magnetic field \cite{6hour}. The ZEFOZ magnetic field of 1.261 Tesla is in the direction $[-0.5914, 0.5239, 0.6130]$ in our laboratory system (and [-0.5600,0.5073,0.6550] in the $[D_1,D_2,b]$). The frequency of the transition $\ket{-3/2}_g\leftrightarrow\ket{+3/2}_g$ at 12.46 MHz is extremely insensitive to magnetic field perturbations.}
\end{figure}

A straightforward application of the results is to predict the hyperfine structures for Eu$^{3+}$ in a critical magnetic field. Zero first order Zeeman (ZEFOZ) point is a magnetic field in which the frequency of the considered hyperfine transition is extremely insensitive to magnetic field perturbations. Since the magnetic field noise from the nuclear spin baths are the primary source of decoherence for Eu$^{3+}$ in Y$_2$SiO$_5$, the coherence time of the considered transition can be extended at ZEFOZ point. Here we take the ZEFOZ magnetic field employed to realize six-hour coherence time \cite{6hour} as an example. The excited-state hyperfine structure is calculated in this field and shown in Fig. \ref{structure}. The ground-state structure is also presented, as a reference.


As a matter of fact, hyperfine structures in any magnetic fields can be predicted based on the fitted spin Hamiltonians. With known energy level structures, optical pumping squences can be designed for the current system in any other ZEFOZ magnetic fields. A three-level system can be well selected within the inhomogeneous broadening and tailored into suitable absorption shape through optical pumping for quantum  memory protocols such as electromagnetic induced transparency (EIT) \cite{eitexp}, controlled reversible inhomogeneous broadening (CRIB) \cite{cribexp} and atomic frequency comb (AFC) \cite{afcexp}.

\section{Conclusion}
By using Raman heterodyne detection of NMR, the hyperfine interactions in the excited state $^5$D$_0$ of $^{151}$Eu$^{3+}$ at site 1 in Y$_2$SiO$_5$ are characterized with the effective spin Hamiltonians. In order to determine how the two $C_2$-symmetric excited states are related to the ground state, we also performed a spectral hole burning experiment in a special magnetic field to eliminate the ambiguity. In our experiments, Raman heterodyne detection shows significant advantages such as vibration-tolerance, high resolution and easiness for implementation and spectral analysis when compared with pure optical methods. The results can find use for selecting a three-level system through optical pumping for quantum memory protocols.




We note that similar results are obtained with optical free induction decay measurements by an independent group \cite{euysoexcitedgisin}.

\section*{Acknowledgements}
This work was supported by the National Key R\&D Program of China (No. 2017YFA0304100, 2016YFA0302700), the National Natural Science Foundation of China (Nos. 61327901, 11774331, 11774335, 11504362, 11654002), Key Research Program of Frontier Sciences, CAS (No. QYZDY-SSW-SLH003), and the Fundamental Research Funds for the Central Universities (Nos. WK2470000023, WK2470000026).


\section*{References}
\begin{thebibliography}{10}
\expandafter\ifx\csname url\endcsname\relax
  \def\url#1{\texttt{#1}}\fi
\expandafter\ifx\csname urlprefix\endcsname\relax\def\urlprefix{URL }\fi
\expandafter\ifx\csname href\endcsname\relax
  \def\href#1#2{#2} \def\path#1{#1}\fi

\bibitem{repeater}
H.~J. Briegel, W.~Dur, J.~I. Cirac, P.~Zoller, Quantum repeaters: The role of
  imperfect local operations in quantum communication, Phys. Rev. Lett. 81
  (1998) 5932--5935.

\bibitem{atomensembleqm}
M.~D. Lukin, Colloquium: Trapping and manipulating photon states in atomic
  ensembles, Rev. Mod. Phys. 75 (2003) 457--472.

\bibitem{atomensembleqm1}
K.~S. Choi, H.~Deng, J.~Laurat, H.~J. Kimble, Mapping photonic entanglement
  into and out of a quantum memory, Nature 452 (2008) 67--71.

\bibitem{atomensembleqm2}
D.~N. Matsukevich, A.~Kuzmich, Quantum state transfer between matter and light,
  Science 306 (2004) 663--666.

\bibitem{singleatomqm}
H.~P. Specht, C.~Nolleke, A.~Reiserer, M.~Uphoff, E.~Figueroa, S.~Ritter,
  G.~Rempe, A single-atom quantum memory, Nature 473 (2011) 190--193.

\bibitem{moleensembleqm}
P.~Rabl, D.~DeMille, J.~M. Doyle, M.~D. Lukin, R.~J. Schoelkopf, P.~Zoller,
  Hybrid quantum processors: molecular ensembles as quantum memory for solid
  state circuits, Phys. Rev. Lett. 97 (2006) 033003.

\bibitem{efficientqm}
M.~P. Hedges, J.~J. Longdell, Y.~Li, M.~J. Sellars, Efficient quantum memory
  for light, Nature 465 (2010) 1052--1056.

\bibitem{hifiqm}
Z.-Q. Zhou, W.-B. Lin, M.~Yang, C.-F. Li, G.-C. Guo, Realization of reliable
  solid-state quantum memory for photonic polarization qubit, Phys. Rev. Lett.
  108 (2012) 190505.

\bibitem{oamentangle}
Z.~Q. Zhou, Y.~L. Hua, X.~Liu, G.~Chen, J.~S. Xu, Y.~J. Han, C.~F. Li, G.~C.
  Guo, Quantum storage of three-dimensional orbital-angular-momentum
  entanglement in a crystal, Phys. Rev. Lett. 115 (2015) 070502.

\bibitem{broadbandqm}
E.~Saglamyurek, N.~Sinclair, J.~Jin, J.~A. Slater, D.~Oblak, F.~Bussieres,
  M.~George, R.~Ricken, W.~Sohler, W.~Tittel, Broadband waveguide quantum
  memory for entangled photons, Nature 469 (2011) 512--515.

\bibitem{longlifeqm}
G.~Heinze, C.~Hubrich, T.~Halfmann, Stopped light and image storage by
  electromagnetically induced transparency up to the regime of one minute,
  Phys. Rev. Lett. 111 (2013) 033601.

\bibitem{EuYSOspectra}
F.~Konz, Y.~Sun, C.~W. Thiel, R.~L. Cone, R.~W. Equall, R.~L. Hutcheson, R.~M.
  Macfarlane, Temperature and concentration dependence of optical dephasing,
  spectral-hole lifetime, and anisotropic absorption in
  {Eu$^{3+}$:Y$_2$SiO$_5$}, Phys. Rev. B 68 (2003) 085109.

\bibitem{6hour}
M.~Zhong, M.~P. Hedges, R.~L. Ahlefeldt, J.~G. Bartholomew, S.~E. Beavan, S.~M.
  Wittig, J.~J. Longdell, M.~J. Sellars, Optically addressable nuclear spins in
  a solid with a six-hour coherence time, Nature 517 (2015) 177--180.

\bibitem{afcexp}
M.~Afzelius, I.~Usmani, A.~Amari, B.~Lauritzen, A.~Walther, C.~Simon,
  N.~Sangouard, J.~Minar, H.~de~Riedmatten, N.~Gisin, S.~Kroll, Demonstration
  of atomic frequency comb memory for light with spin-wave storage, Phys. Rev.
  Lett. 104 (2010) 040503.

\bibitem{cribexp}
M.~Nilsson, S.~Kroll, Solid state quantum memory using complete absorption and
  re-emission of photons by tailored and externally controlled inhomogeneous
  absorption profiles, Opt. Commun. 247 (2005) 393--403.

\bibitem{eitexp}
A.~V. Turukhin, V.~S. Sudarshanam, M.~S. Shahriar, J.~A. Musser, B.~S. Ham,
  P.~R. Hemmer, Observation of ultraslow and stored light pulses in a solid,
  Phys. Rev. Lett. 88 (2002) 023602.

\bibitem{SHBNdYVO}
S.~R. Hastings-Simon, M.~Afzelius, J.~Minar, M.~U. Staudt, B.~Lauritzen,
  H.~de~Riedmatten, N.~Gisin, A.~Amari, A.~Walther, S.~Kroll, E.~Cavalli,
  M.~Bettinelli, Spectral hole-burning spectroscopy in {Nd$^{3+}$:YVO$_4$},
  Phys. Rev. B 77 (2008) 125111.

\bibitem{REbookMacSHB}
A.~A. Kaplyanskii, R.~M. Macfarlane (Eds.), Spectroscopy of Solids Containing
  Rare Earth Ions, North-Holland, Amsterdam, 1987, Ch.~1, pp. 62--67.

\bibitem{ramanprb}
N.~C. Wong, E.~S. Kintzer, J.~Mlynek, R.~G. DeVoe, R.~G. Brewer, Raman
  heterodyne detection of nuclear magnetic resonance, Phys. Rev. B 28 (1983)
  4993--5010.

\bibitem{ramanprl}
J.~Mlynek, N.~C. Wong, R.~G. DeVoe, E.~S. Kintzer, R.~G. Brewer, Raman
  heterodyne detection of nuclear magnetic resonance, Phys. Rev. Lett. 50
  (1983) 993--996.

\bibitem{groundeu}
J.~J. Longdell, A.~L. Alexander, M.~J. Sellars, Characterization of the
  hyperfine interaction in europium-doped yttrium orthosilicate and europium
  chloride hexahydrate, Phys. Rev. B 74 (2006) 195101.

\bibitem{groundpr}
J.~J. Longdell, M.~J. Sellars, N.~B. Manson, Hyperfine interaction in ground
  and excited states of praseodymium-doped yttrium orthosilicate, Phys. Rev. B
  66 (2002) 035101.

\bibitem{REbookLiu}
G.~Liu, B.~Jacquier, Spectroscopic Properties of Rare Earths in Optical
  Materials, Springer, Berlin, 2005, Ch.~1, p.~77.

\bibitem{REbookMacHam}
A.~A. Kaplyanskii, R.~M. Macfarlane (Eds.), Spectroscopy of Solids Containing
  Rare Earth Ions, North-Holland, Amsterdam, 1987, Ch.~3, pp. 85--90.

\bibitem{SHBEuCl3}
J.~P.~D. Martin, M.~J. Sellars, P.~Tuthill, N.~B. Manson, G.~Pryde, T.~Dyke,
  Resolved isotopic energy shift and hole burning in {EuCl$_3\cdotp$6H$_2$O},
  J. Lumin. 78 (1998) 19--24.

\bibitem{EuYSOwavelength2013}
M.~J. Thorpe, D.~R. Leibrandt, T.~Rosenband, Shifts of optical frequency
  references based on spectral-hole burning in {Eu$^{3+}$:Y$_2$SiO$_5$}, New J.
  Phys. 15 (2013) 033006.

\bibitem{EuYSOwavelength1991}
R.~Yano, M.~Mitsunaga, N.~Uesugi, Ultralong optical dephasing time in
  {Eu$^{3+}$:Y$_2$SiO$_5$}, Opt. Lett. 16 (1991) 1884--1886.

\bibitem{EuYSOnonlinear}
R.~Yano, M.~Mitsunaga, N.~Uesugi, Nonlinear laser spectroscopy of
  {Eu$^{3+}$:Y$_2$SiO$_5$} and its application to time-domain optical memory,
  J. Opt. Soc. Am. B 9 (1992) 992--997.

\bibitem{c2angle}
C.~J. Taylor, D.~J. Kriegman, Minimization on the {Lie} group {SO}(3) and
  related manifolds, Tech. Rep. 9405, Yale University (1994).

\bibitem{simuanneal}
S.~Kirkpatrick, J.~Gelatt, C.~D., M.~P. Vecchi, Optimization by simulated
  annealing, Science 220 (1983) 671--680.

\bibitem{151EuYSOmemory}
P.~Jobez, C.~Laplane, N.~Timoney, N.~Gisin, A.~Ferrier, P.~Goldner,
  M.~Afzelius, Coherent spin control at the quantum level in an ensemble-based
  optical memory, Phys. Rev. Lett. 114 (2015) 230502.

\bibitem{EuYSOabsorption}
A.~Ferrier, B.~Tumino, P.~Goldner, Variations in the oscillator strength of the
  $^7${F}$_0$ $\rightarrow$ $^5${D}$_0$ transition in {Eu$^{3+}$:Y$_2$SiO$_5$}
  single crystals, J. Lumin. 170 (2016) 406--410.

\bibitem{SHB}
M.~Nilsson, L.~Rippe, S.~Kroll, R.~Klieber, D.~Suter, Hole-burning techniques
  for isolation and study of individual hyperfine transitions in
  inhomogeneously broadened solids demonstrated in {Pr$^{3+}$:Y$_2$SiO$_5$},
  Phys. Rev. B 70 (2004) 214116.

\bibitem{SHBEuYSO}
B.~Lauritzen, N.~Timoney, N.~Gisin, M.~Afzelius, H.~de~Riedmatten, Y.~Sun,
  R.~M. Macfarlane, R.~L. Cone, Spectroscopic investigations of
  {Eu$^{3+}$:Y$_2$SiO$_5$} for quantum memory applications, Phys. Rev. B 85
  (2012) 115111.

\bibitem{euysoexcitedgisin}
E.~Z. Cruzeiro, J.~Etesse, A.~Tiranov, P.-A. Bourdel, F.~Fröwis, P.~Goldner,
  N.~Gisin, M.~Afzelius, Characterization of the hyperfine interaction of the
  excited 5d0 state of {Eu$^{3+}$:Y$_2$SiO$_5$}, arXiv:1710.07591.

\end{thebibliography}
\end{document}